\documentclass{emulateapj}

\def\be{\begin{equation}}
\def\ee{\end{equation}}
\def\kms{\rm km s^{-1}}
\def\tauq{\tau_{\rm Q}}
\def\mpc{\rm Mpc}
\def\yr{\rm yr}
\def\ergshz{\rm erg~s^{-1}~Hz^{-1}}

\shorttitle{Probing QSO properties through proximity effect}
\shortauthors{Lu \& Yu}

\begin{document}
\title{On probing the properties of QSOs through their proximity effects on
the intergalactic medium
}
\author{Youjun Lu$^1$ \& Qingjuan Yu$^2$}
\affil{
~$^1$ National Astronomical Observatories, Chinese Academy of Sciences,
Beijing, 100012, China; luyj@nao.cas.cn\\
~$^2$ Kavli Institute for Astronomy and Astrophysics, Peking University,
Beijing, 100891, China; yuqj@pku.edu.cn
}

\begin{abstract}

The proximity effect (PE) of QSOs is believed to be useful in constraining the
QSO lifetime. Observations on the PE so far, however, give apparently
contradictory results -- some are consistent with a long QSO lifetime ($\ga$ a
few $10^7$~yr), but others appear to be only consistent with a short QSO
lifetime $\la 10^6$~yr. In this paper, we show that this apparent contradiction
may be solved by simultaneously taking into account both the effect due to the
density enhancement in the QSO near zones and that due to the obscuration of
the tori associated with the QSOs, using a large number of Monte-Carlo
generated synthetic Ly$\alpha$ forest spectra. We demonstrate that the QSO
properties and environment can be constrained simultaneously by the transverse
PE and the line of sight PE of bright type 1 QSOs together. The current
available measurements on the PEs of type 1 QSOs suggest that (1) the density
is significantly enhanced in the vicinity of the QSOs; (2) the QSO lifetime is
consistent with being as large as a few $10^7$~yr and a substantially shorter
lifetime (e.g., $\la 10^6$~yr) is not required; and (3) the half opening angle
of the tori associated with QSOs is $\sim 60\degr$, consistent with some other
independent estimates. Our simulations also show that the TPE of type 2 QSOs
can be significantly different from that of type 1 QSOs, which may be useful
to put further constraints on the QSO properties and the QSO environment.

\end{abstract}


\section{Introduction }\label{sec:intro}

The enormous UV radiation from a QSO can significantly alter the ionization
state of its surrounding intergalactic medium (IGM).  The Ly$\alpha$ absorption
clouds near the QSO are more ionized than average, and the optical depth of
Ly$\alpha$ photons due to the absorbers was expected to decrease as the
absorption redshift approaches that of the QSO, which is named as the
``proximity effect'' (PE) \citep{Carswell87,BDO88}. The significance of the PE
of a QSO is mainly determined by the following factors: (1) the strength of the
metagalactic UV ionizing background (UVB); (2) the intrinsic properties of the
QSO, including its luminosity, its age (or its detailed luminosity evolution),
and the anisotropic feature in its UV radiation; and (3) the overdense
environment of the IGM surrounding the QSO.  Therefore, the PE is useful not
only to measure the UVB \citep[e.g.,][]{BDO88}, but also to constrain the
lifetime/luminosity evolution of QSOs and the density enhancement in its near
zones \citep[e.g.,][]{Jakobsen03,Croft04,Schirber04, Adelberger04, Rollinde05,
Guimaraes07,KT08,GSP08}.

The PE along the line of sight to a QSO (LOSPE) has been clearly detected
through the high-resolution Ly$\alpha$ absorption spectra of a combined QSO
sample or some individual QSOs, which is widely used to estimate the strength
of the UVB \citep[e.g.,][]{Scott00,Dall08} and also the density enhancement of
the IGM surrounding the QSO \citep[e.g.,][]{Rollinde05,Guimaraes07}. For
example, \citet{Dall08} and \citet{FGetal08b} find that the UVB measured from
the LOSPE (after including the effect of the overdense environment of QSOs) is
consistent with those estimates obtained from other methods. The LOSPE is
affected directly by the observed QSO luminosity, but almost irrelevant to the
QSO's past luminosity evolution (including luminosity variation) and the age of
its nuclear activity. In principle, the PE may be also affected by fluctuations
in the QSO luminosity on short timescales $\la2\times 10^4\yr$, over which the
perturbed ionization state of the highly ionized IGM can get back to the
photoionization equilibrium.  However, the QSO variability studies give the
average lifetime of an individual episode of QSO activity $\ga10^4\yr$
\citep[e.g.,][]{MS03,Martini04}.

The PE in the region other than along the line of sight to the
QSO (i.e., the transverse PE or TPE) can be measured through
the spectra of background QSOs (bgQSOs) or other bright UV sources
whose light passed through the highly ionized region surrounding the foreground
QSOs (fgQSOs). Unlike the LOSPE, the TPE should be sensitive to the age of the
fgQSO, as a positive detection of this effect requires that the light travel
time from the fgQSO to the line of sight (LOS) to the bgQSO is smaller
than the age of the fgQSO. Hence, the TPE has long been thought to be helpful
in constraining the QSO lifetime \citep[e.g.,][]{BDO88,Croft04,
Schirber04,Adelberger04,KT08}. However, so far the results from observations
on the significance of the TPE appear ambiguous. 
For example, many authors tried to search for the reduction in
the optical depth or the number of HI Ly$\alpha$ forest lines caused by the TPE
through QSO pairs with small transverse separation (typically $\la$ a few Mpc),
but they found no evidence of this reduction and suggested that the QSO
lifetime $\la10^6\yr$ \citep[e.g.,][]{FS95,CF98,Croft04, Schirber04,KT08}.
However, several authors have reported the detections of the TPE through the
observations on the He~{\small II}, rather than H~{\small I}, Ly$\alpha$ forest
or metal absorption lines \citep[e.g.,][]{Jakobsen03,WW06,Worseck07,GSP08}.
Those detections suggest that the QSO lifetime should be $\ga10^7\yr$.
Currently it is still unclear whether that apparent contradiction is due to
some selection effects in observations or the complications
introduced by other factors, such as, the density enhancement in the QSO
near zones or the anisotropic radiation from QSOs.

In this paper, we study the LOSPE and the TPE of QSOs, with taking into account
the density enhancement in the QSO proximity region, their anisotropic UV
radiation, and their lifetime; and the purpose is not only to solve the
apparent contradiction among the observational constraints on the QSO lifetime
from current TPE observations, but also to demonstrate that the LOSPE and the
TPE combine to put constraints on both the QSO lifetime and the opening angle
of torus associated with QSOs simultaneously. The paper is organized as
follows.  In Section~\ref{sec:Geom}, we illustrate the geometry of the
proximity region of a QSO in the observer's rest frame that can be affected by
UV photons emitted from the QSO. Accounting the PE of (fg)QSOs, we use a
Monte-Carlo method to generate a large number of Ly$\alpha$ forest spectra of
fgQSOs and bgQSOs as described in Section~\ref{sec:MCsimu}, based on some
statistical distributions of the HI column density, the Doppler factor of
Ly$\alpha$ absorption lines, and the redshift distribution of the number
density of these lines \citep[see][]{Dall08,Dall09}. With these synthetic
Ly$\alpha$ forest spectra, we illustrate both the optical depth decrease due to
the LOSPE and the TPE of QSOs and then compare them with current observational
results on type 1 QSOs in Section~\ref{sec:results}. Considering that future
X-ray observations potentially discover a large number of type 2 QSOs, the TPE
of type 2 QSOs is also investigated in Section~\ref{sec:results}.  Our
simulation results demonstrate that the density enhancement near the QSOs, the
QSO lifetime, and the opening angle of the tori associated with the QSOs can be
simultaneously constrained through measurements on both the LOSPE and the TPE.
Discussions and conclusions are given in Section~\ref{sec:conclusion}.

In the paper, We adopt the Hubble constant $H_0=71\kms$ and the cosmological
parameters $(\Omega_{\rm m},\Omega_{\Lambda})=(0.27,0.73)$ \citep{Komatsu09}. 

\section{Geometry of the proximity region of a QSO}\label{sec:Geom}

\begin{figure*}
   \includegraphics[width=6.5in]{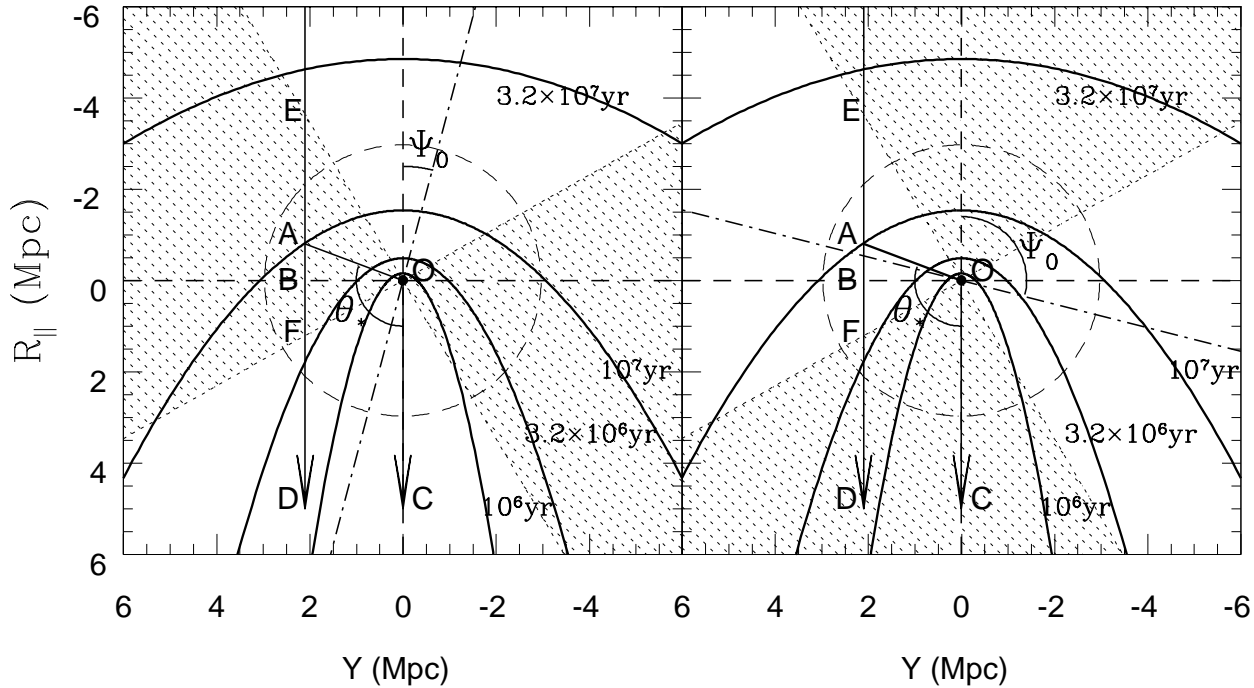}
   \caption{Schematic diagram for the proximity region of a fgQSO that may be
affected by the UV photons emitted from it. The fgQSO is at point O, and the
distant observer is located in the direction OC.  The ED is the direction from
a background bright UV source or QSO to the observer.  The left and right
panels illustrate the cases for a type 1 fgQSO and a type 2 fgQSO,
respectively. The intrinsic QSO UV radiation is assumed to be isotropic, but it
may be shielded in some directions by a dusty torus as shown by the shaded area
in both panels.  The half opening angle of the torus is $\Theta_0=45\degr$, and
the offset of the torus axis (the dot-dashed lines) from the direction OC is
$\Psi_0=15\degr $ in the left panel and $105\degr  $ in the right panel,
respectively.  The shaded region is rotationally symmetric around the torus
axis. For illustration purpose, the LOS to the bgQSO is assumed to be
specifically on the plane of the torus axis and the LOS to the fgQSO ( see
Appendix for discussion on more general cases); and it has a physical distance
of $R_\perp=|{\rm OB}|$ from the fgQSO and interacts with the shaded region at
points E and F. The thick solid curves represent the envelope surfaces of the
proximity regions that can be affected by the fgQSO radiation for the QSO age
of $3.2\times10^7\yr$, $10^7\yr$, $3.2\times10^6\yr$, and $10^6\yr$ (from top
to bottom), respectively.  These surfaces are rotationally symmetric around the
LOS to the fgQSO.  The dashed circle represents the surface on which the
photoionization rate enhancement due to the central QSO is
$\omega\equiv\Gamma_{\rm QSO}/\Gamma_{\rm UVB}=1$. Here the QSO is set to be at
redshift $z_{\rm fg}=2$ with a Lyman limit luminosity of $10^{30.5}{\rm erg
s^{-1} Hz^{-1}}$, and the photoionization rate due to the UVB $\Gamma_{\rm
UVB}=10^{-12}{\rm s^{-1}}$.   }
\label{fig:f1} \end{figure*}

In this section, we illustrate how the geometry of the proximity region of a
fgQSO in the distant observer's frame is affected by the age of the QSO and the
anisotropic feature of its radiation. In general, we describe the region
near a fgQSO by the physical distance of a point from the QSO $R$ and
its polar angle $\theta$ measured from the observer's LOS (i.e., OC in
Figure~\ref{fig:f1}).

From Ly$\alpha$ forest spectra, the distant observer deduces the proximity
region which can be affected by the fgQSO radiation through the light passing
by the region. Considering the finite speed of the light, the combination of
the light traveling time from the affected region to the observer and the time
of the fgQSO light traveling from its location to its affected region should
not be longer than the light traveling time from the fgQSO to the observer by
the QSO age. That is, in the distant observer's rest frame, photons emitted
from the QSO can reach a physical distance of at most $R_*(\theta_*)=|{\rm
OA}|$ from the QSO as shown in Figure~\ref{fig:f1}, and
$R_*(1-\cos\theta_*)=c\tauq$, where $\theta_*$ is the angle of OA from the
observer's LOS OC, $\tauq$ is the QSO age, and $c$ is the speed of light. The
$R_*(\theta_*)$ gives the apparent time-delay envelope within which the
ionization state of (hydrogen) atoms can be possibly affected by UV radiation
from the QSO.

Considering a bgQSO of which the LOS to the observer is separated from the
fgQSO by a transverse proper distance of $R_{\perp}$, the intersection point of
the bgQSO LOS with the time-delay envelope (point A in Figure~\ref{fig:f1}) is
given by $\theta_*=2\arctan( c\tauq/R_{\perp})$. The Ly$\alpha$ forest in the
spectrum of the bgQSO may be then affected by the fgQSO at observational
wavelength $\lambda\leq1216{\rm \AA}(1+z_A)$, where $z_A$ is the redshift of
the interaction point A.  We have $z_A\simeq z_{\rm fg}-(1+z_{\rm fg})E(z_{\rm
fg})R_{\parallel,\rm A}/c$, where $z_{\rm fg}$ is the redshift of the fgQSO,
$R_{\parallel,\rm A}=\frac{\cos\theta_*}{1- \cos\theta_*}c\tauq$ is the the
physical distance from A to B shown in Figure~\ref{fig:f1}, point B is located
on the LOS to the bgQSO and has the same distance/redshift to the observer as
the fgQSO, and $E(z)=H_0\sqrt{\Omega_{\rm m}(1+z)^3+\Omega_{\Lambda}}$.  The
photoionization enhancement at a particular point on the LOS to the bgQSO due
to the fgQSO can be characterized by the ratio of $\omega\equiv\Gamma_{\rm
QSO}/\Gamma_{\rm UVB}$, where $\Gamma_{\rm QSO}$ is the photoionization rate
due to the QSO UV radiation and $\Gamma_{\rm UVB}$ is the photoionization rate
due to the cosmic UVB. The latest constraints on the cosmic UVB show that
$\Gamma_{\rm UVB}$ is roughly a constant over redshift from 2 to 4 (e.g., $\sim
0.5\times 10^{-12}{\rm s}^{-1}$ in \citealt{FGetal08a}, $\sim 1-1.3\times 10^{-12}
{\rm s}^{-1}$ in \citealt{Bolton05}, 
see also \citealt{HM96,Rauch97,MM01,MW04,Tytler04,Kirkman05}). 
In this paper, we choose $\Gamma_{\rm UVB}\simeq 10^{-12}{\rm s}^{-1}$. The $\Gamma_{\rm
QSO}$ at a distance $R$ from the fgQSO is $\int^{\infty}_{\nu_0} \frac{L_{\rm
QSO,\nu}(t)}{4\pi R^2}\frac{\sigma_{\rm HI}(\nu)}{h\nu}d\nu$, where $L_{\rm
QSO,\nu}(t)$ is the luminosity of the QSO per unit frequency at its intrinsic
frequency $\nu$ emitted at time $t=\tauq-R(1-\cos\theta)/c$ since its nuclear
activity is triggered, $\nu_0$ is the Lyman limit frequency, $\sigma_{\rm
HI}(\nu)\propto \nu^{-3}$ is the HI photoionization cross-section
\citep{Osterbrock}, and $R$ can be approximated as the corresponding luminosity
distance as long as it is small enough. In this paper, we assume $L_{\rm
QSO,\nu}\propto (\nu/\nu_0)^{-0.5}$.  Due to the time delay, we have
$\Gamma_{\rm QSO}=0$ and $\omega=0$ at those points with $R>R_*(\theta)$.

In the AGN unification model \citep[e.g.,][]{UP95}, however, the UV radiation
from a QSO can be highly anisotropic depending on whether the torus associated
with the QSO blocks its radiation along the LOS to the observer, and this
geometrical effect results in the observational differences between type 1 and
type 2 QSOs.  The anisotropic radiation of QSOs leads to different ionization
states of its surrounding IGM in different directions and thus we may have
different observational PEs for type 1 and type 2 QSOs as follows.
\begin{itemize}

\item For a type 1 QSO, the UV radiation may be blocked in the transverse
direction and the ionization state of the IGM in that direction may be not
affected by the UV photons from the QSO. We denote the half opening angle of
the dusty torus associated with the fgQSO by $\Theta_0$
($0\degr\le\Theta_0\le90\degr$) and the torus axis offset from the LOS by an
angle of $\Psi_0$ ($0\degr\le\Psi_0<180\degr$; see the left panel of Figure
\ref{fig:f1}).  The intersections of the bgQSO LOS with the light cone confined
by the torus associated with the fgQSO are E and F. Given a transverse distance
$R_{\perp}$, the proper distances from E to B and F to B are given by the two
solutions ($R_{\parallel,\rm E}$, $R_{\parallel,\rm F}$) of the equations
governing the light cone and the cylinder with a radius of $R_{\perp}$ (see
details in Appendix), and correspondingly the redshifts of E and F are given by
$z_E\simeq z_{\rm fg}-(1+z_{\rm fg})E(z_{\rm fg})R_{\parallel,\rm E}/c$ and
$z_F\simeq z_{\rm fg}-(1+z_{\rm fg})E(z_{\rm fg})R_{\parallel,\rm F}/c$,
respectively.  Therefore, the UV photons from the fgQSO can only affect the
spectrum of the bgQSO in the wavelength ranges of $1216{\rm \AA}\min(1+ z_{\rm
E},1+z_{\rm A})\leq\lambda\leq 1216{\rm \AA}(1+z_{\rm A})$ and
$\lambda\leq1216{\rm \AA}\min(1+z_{\rm A},1+z_{\rm F})$ as $z_{\rm E}>z_{\rm
F}$ (see the left panel of Figure~\ref{fig:f1}). Outside of these wavelength
ranges, $\omega=0$.

\item For a
type 2 QSO, the UV radiation is blocked from the view of a distant observer by
a dusty torus. This QSO cannot be seen in the optical-UV band but is detectable
in the hard X-ray. However, the UV radiation from a type 2 QSO may
significantly alter the ionization state of the IGM in the transverse direction
of its proximity region (see the right panel in Figure~\ref{fig:f1}).  For a
type 2 QSO,  the UV photons can affect the Ly$\alpha$ forest spectrum of a
bgQSO in the wavelength range of $1216{\rm \AA}\min(1+z_{\rm E},1+z_{\rm
A})\leq\lambda\leq1216{\rm \AA}\min(1+z_{\rm A},1+z_{\rm F})$ (see the right
panel of Figure~\ref{fig:f1}). For those cases without solutions to the
two equations, the spectrum of the bgQSO is not affected by the PE
with $\omega=0$.
\end{itemize}

We note here that the TPE of type 2 fgQSOs could be the same as that of type 1
fgQSOs if the obscuration, which leads to the classification of type 1 and
type 2 QSOs, is due to isotropically distributed clumpy absorbers, rather than
the uniform torus-like structure.

\section{Monte-Carlo simulations of synthetic spectra of Ly$\alpha$ forests near 
a fgQSO} \label{sec:MCsimu}

In principle, both the LOSPE and TPE can be well understood by detailed 3D
numerical simulations that can successfully reproduce the Ly$\alpha$ forest at
any redshift \citep[e.g.,][]{Croft04,Schirber04,FGetal08b}. Those simulations
with detailed radiative transfer may be able to accommodate all the statistical
and systematic effects and the density enhancement in the vicinity of QSOs, but
they are very time consuming and dependent on the assumptions of the QSO host
dark matter halos. In this paper, we alternatively adopt a Monte Carlo method,
proposed by \citet[][see also \citealt{Dall08,Dall09}]{WW06}, to generate
Ly$\alpha$ forest spectra for a large number of LOSs. This Monte-Carlo method
has been demonstrated to be able to reproduce the Ly$\alpha$ forest spectra of
QSOs similar to observed ones and be useful in measuring the UVB through the
QSO proximity effect \citep{Dall08}.  With detailed statistical consideration
on different factors that may affect the proximity effect of the fgQSOs,
including the QSO age, the anisotropic UV radiation and the density
enhancement, etc. (as discussed in Section~\ref{sec:Geom}), we simulate both
the LOSPE and TPE for a large number of fgQSOs.

The procedure adopted to generate the mock QSO Ly$\alpha$ forest spectra is
based on the observations that each Ly$\alpha$ absorption line with Voigt
profile can be described by the HI column density $N_{\rm HI}$ and the Doppler
parameter $b$ of its corresponding absorption gas and that the comoving number
density of the absorption lines per unit $z$, $N_{\rm HI}$, and $b$, denoted by
$n(z,N_{\rm HI},b)$, can be well described by the following three
distributions: \begin{enumerate}

\item the redshift distribution, approximated by a power-law form with
$n\propto (1+z)^{\gamma}$, where $\gamma=2.13$ \citep[][see also
\citealt{FGetal08c}]{Kim01,Schaye03}; 

\item the HI column density distribution with $n \propto N_{\rm HI}^{-\beta}$,
where the power $\beta\simeq 1.5$ \citep{Kim01}; 

\item the Doppler parameter distribution with $n\propto
b^{-5}\exp[-b_{\sigma}^4/b^4]$ where $b_{\sigma}\simeq 24\kms$ \citep{Kim01}. 

\end{enumerate}

The column densities of the simulated absorbers are limited to be within the
range $10^{12}{\rm cm}^{-2}<N_{\rm HI}<10^{18}{\rm cm}^{-2}$ and the Doppler
parameters within $10\kms<b<100\kms$. Those three distributions are assumed to
be independent one another. For the purpose of this paper, we populate
absorbers in each line of sight according to the above distributions and
generate mock samples of the Ly$\alpha$ forest spectra. Gaussian noise is
added to the simulated Ly$\alpha$ forest spectra and the spectral resolution is
also considered in order to match the signal-to-noise level of those observed
samples discussed in Section~4. For simplicity, the emission lines are not
added to the spectra of mock QSOs.  Note that the adopted Monte-Carlo method
naturally includes the Poisson variance of the number of absorption lines in
the near zone of fgQSOs \cite[e.g.,][]{Dall08}.

In the QSO near zones, the mass density may be significantly enhanced relative
to the cosmic average density, i.e., $\left<\Delta(R)\right>\equiv
\left<\rho(R)\right>/\bar{\rho}>1$, where $\left<\Delta(R)\right>$ and
$\left<\rho(R)\right>$ is the mass overdensity and the mass density at a
distance $R$ from central QSOs averaged over a large sample of QSOs with
similar properties, respectively, and $\bar{\rho}$ is the cosmic average
density. A simple photoionization equilibrium model can give the neutral
hydrogen density of an absorber $n_{\rm HI} \propto \delta^{2-0.7\eta}$, where
$\delta$ is the mass overdensity of the absorber $\delta$ and $\eta\sim 0.62$
is the index of the power-law temperature-density relation for the
low-temperature IGM \citep[e.g.,][]{FGetal08b,Schaye01,HG97}.
Assuming that the column density
distribution of Ly$\alpha$ absorbers in the QSO near zones (with mean density
enhancement factor $\left<\Delta(R)\right>$) also follows a power law $\propto
N_{\rm HI}^{-\beta}$ with the same slope $\beta$ as that for the cosmic average
(item 2 listed above), the column density of each absorber
generated from the above distribution (item 2) should then be replaced by
$N_{\rm HI}\left<\Delta(R)\right>^{2-0.7\eta}/(1+\omega)$, considering of both
the effects of the enhanced density and UV ionizing flux in QSO
near zones \citep[cf., see Equation 30 in][]{FGetal08b}. Note that the effect
of that replacement is equivalent to the effect of increasing the number of
absorbers proportionally.

As illustrated in Figure \ref{fig:f1}, different choices of the fgQSO age, the
half opening angle $\Theta_0$, and the offset angle $\Psi_0$ of the torus
associated with the fgQSO will affect the proximity region that the fgQSO can
illuminate. In our simulations, we set seven different values for the lifetime
of QSOs $\tau_{\rm lt}$, and the QSO age $\tauq$ is either fixed or randomly
chosen over the range $(0,\tau_{\rm lt})$. We choose six sets of values for the
opening angle of the torus, i.e., $\Theta_0=0\degr, 30\degr, 45\degr, 60\degr,
72\degr$, and $89\degr$. The $\Theta_0=0\degr$ corresponds to the cases of no
ionization PE from the central QSOs, and the $\Theta_0=89\degr$ corresponds
closely to no obscuration to the UV radiation from the fgQSOs.  The other
specific values of $\Theta_0$, i.e., $30\degr$, $45\degr$, $60\degr$, and
$72\degr$, are chosen as their corresponding abundance ratios of type 2 to type
1 QSOs (6.5, 2.4, 1, and 0.4, respectively) are roughly in the range estimated
by observations \citep[e.g.,][]{TU06,Gilli07,Treister10}.  We also randomly
choose $\Psi_0$ within the range from $\Theta_0$ to $180\degr-\Theta_0$ for
type 2 QSOs, and other range for type 1 QSOs.

According to the above settings, we first simulate the Ly$\alpha$ forest
spectra of QSOs affected by its own UV radiation to account for the LOSPE.
In order to compare with the observational results on the LOSPE (and the TPE
later) by \citet{KT08} (hereafter KT08) obtained from a sample of 130 QSO
pairs, the observational redshift of the fgQSOs and its luminosities at the
Lyman limit frequency in our simulation are set to be $z_{\rm fg}=2$ and
$L_{\nu_0}=5\times 10^{30}\ergshz$, which are roughly the mean redshift and
luminosity of the sample in KT08. Note that most of the QSOs in KT08 sample are
in the redshift range $z\sim 1.8-2.6$, and the redshift dependence of the
optical depth has been corrected in KT08. As the actual halos hosting these
fgQSOs and the overdensity distribution surrounding them are not well known, we
only consider the average effects of the sample in this paper, not going to
the detailed redshift and luminosity distributions of QSOs as shown in the KT08
sample. We first create 100
realizations of 130 independent synthetic spectra to check whether the sample
variance could be significant in the interpretation of the observational
results. And then we also create larger mock samples with 500 independent
synthetic Ly$\alpha$ forest spectra to check whether the QSO properties can be
extracted effectively from a sample with 500 spectra or more. Similar to KT08,
the amount of absorption in each spectrum of the mock samples is quantified by
$DA\equiv1-F/C$, where $F$ is the flux and $C$ is the continuum level, and a
uniform flux decrement due to the metal absorption $DA_{\rm metal}=0.025$ is
also added to each spectrum at wavelength higher than the Ly$\alpha$ emission
line. We also simulate the Ly$\alpha$ forest spectra of bgQSOs, which are
affected by the UV radiation from fgQSOs, to account for the TPE. The
transverse distances of fgQSOs from the LOS to their bgQSOs $R_{\perp}$ are
randomly chosen from $0.1$~Mpc to $3$~Mpc, within the range of the
observational sample in KT08. We also create 500 independent synthetic spectra
for each set of parameters ($\tau_{\rm age},\Theta_0$) for both type 1 and type
2 fgQSOs, respectively. For each mock sample, we then obtain the average flux
decrement $\langle DA \rangle$ of the 500 spectra.  Comparison of the simulated
LOSPE and TPE with the observations may thus put some constraints on the
intrinsic properties of QSOs.

Similarly, we create 500, if not otherwise specified, independent synthetic
spectra of a QSO/bgQSO, which are affected by either its own UV radiation or
the UV radiation from a fgQSO at $z_{\rm fg}=4$ with a mean Lyman limit
luminosity of $L_{\nu_0}= 8\times 10^{31}\ergshz$, for each set of
parameters ($\tau_{\rm lt}$, $\Theta_0$). The LOSPE obtained from the set of
synthetic Ly$\alpha$ forest spectra can be used to compare with that obtained
by \citet{Guimaraes07} for a sample of QSOs with similar redshift and mean
Lyman limit luminosity.  Similar to \citet{Guimaraes07}, the absorption here is
quantified by the median of the optical depths $\tau\equiv-\ln(F/C)$ of all the
LOSs.  With some constraints obtained from the LOSPE observations by
\citet{Guimaraes07}, we also obtain the expected TPE of type 1 and type 2
fgQSOs from the mock samples.

\section{Simulation results}\label{sec:results}
 
We illustrate the simulation results on both the LOSPE and TPE for both the
samples with relatively low luminosities at low redshift ($z\sim 2$ and
$L_{\nu_0}\sim 5\times10^{30}\ergshz$; Section~\ref{sec:low}) and the samples
with high luminosities at high redshift ($z\sim 4$ and $L_{\nu_0}\sim 8\times
10^{31}\ergshz$; Section~\ref{sec:high}). In order to compare with those
observations by KT08 and \citet{Guimaraes07}, the spectra are averaged over
resolution of $1{\rm \AA}$ and $0.3{\rm \AA}$, and the signal to noise ratio
are set to be $11.2$ and $25$ for the low-luminosity and high-luminosity
samples similar to those in KT08 and \citet{Guimaraes07}, respectively.  Along
with the detailed presentation on how the detected PEs can be affected by the
QSO lifetime, the geometric structure of the torus, and the QSO environment
below, the effect of different QSO luminosities is also revealed.

\subsection{Low-luminosity samples at $z\sim2$ }\label{sec:low}

\subsubsection{The LOSPE of type 1 fgQSOs}
\label{sec:sublospe1}

\begin{figure}
\includegraphics[width=84mm]{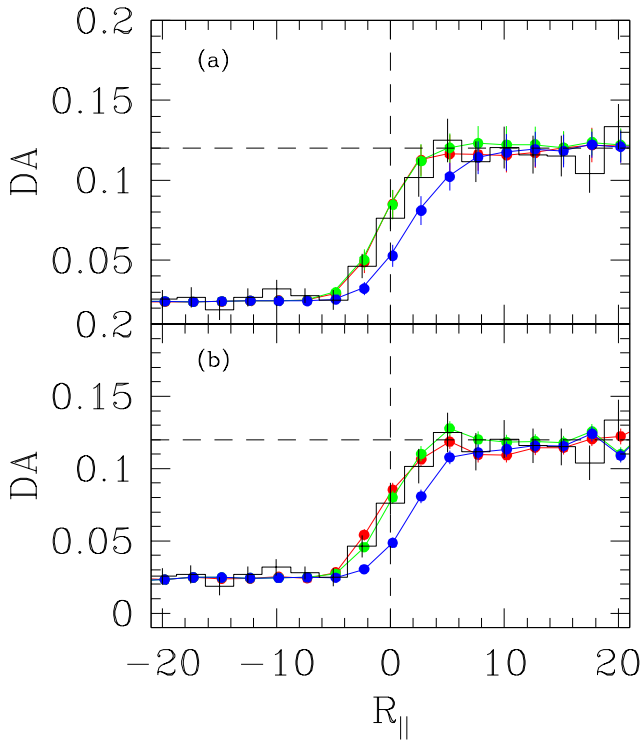}
\caption{The line of sight proximity effect. The histogram in panel a (or b)
shows the measured flux decrement $DA$ at a proper distance $R_{\parallel}$
along the LOS from the central QSO, averaged over an observational sample of
130 QSOs with redshift $\sim 2$ and Lyman limit luminosity
$L_{\nu_0}\sim5\times 10^{30}\ergshz$ (adopted from KT08).  The color lines and
points in panel (a) represent the mean $DA$ obtained from 100 realizations of a mock
sample with 130 synthetic Ly$\alpha$ forest spectra, where the error bar for
each bin is the standard deviation of the sample, while in panel (b) they represent
that obtained from mock samples with 500 synthetic Ly$\alpha$ forest spectra, 
where the error bar is the standard error on the mean.  The blue 
line and solid circles show the simulated $DA$ due to the PE of the central QSOs
if the average density in the QSO near zones is similar to the cosmic average;
the green line and solid circles show the simulated $DA$ if there is no proximity
effect due to the UV radiation from the central QSO and no density enhancement
in the QSO near zones; and the red line and solid circles show the simulated $DA$ 
if the density is enhanced in the QSO near zones effectively by a factor of
$\left<\Delta(R)\right>$ (see details in Section~\ref{sec:sublospe1}). All the
circle points are centered on the same bins of $R_{\parallel}$ as the
observational data but have been slightly offset to the right for graphical
clarity. 
} \label{fig:f2}
\end{figure}

Figure \ref{fig:f2} shows the average absorption near QSOs/fgQSOs along their
LOS with the origin corresponding to their redshift. The histogram with error
bars shows the $DA$ measured by KT08. The color lines and symbols in panels (a)
and (b) show the results obtained from 100 realizations of a mock sample with 130
synthetic Ly$\alpha$ forest spectra and the results from a mock sample with 500
synthetic spectra, respectively. The value plotted for each bin is the mean
value of $DA$ from the Ly$\alpha$ forest spectra, and the error bar is taken to
be standard deviation of the 100 realizations in panel (a) and the
standard error of the mean in panel (b), respectively.  The color lines and
symbols show the expected $DA$ estimated by assuming the following several
cases: (1) on average the density in the QSO near zones is not enhanced (blue
line and points), i.e., $\left<\Delta(R)\right>=1$; or (2) there is no UV
radiation from the central QSOs and on average the density near the QSO is not
enhanced, i.e., $\omega=0$ and $\left<\Delta(R)\right>=1$ (green line and
points); or (3) there is a density enhancement near the QSO described by
$\left<\Delta(R)\right>\sim 1+C_0(R/C_R)^{-p} \exp[-(R/C_R)^q]$ with $C_0=2.2$,
$C_R=3.0\mpc$, $p=0.7$, and $q=2.0$ (red line and points)\footnote{Note that
this assumed form of $\left<\Delta(R)\right>$ follows the tendency of
approaching 1 at sufficiently distant regions and increasing with the decreased
$R$. Here $\left<\Delta(R)\right>$ can be taken as the effective density
enhancement factor but not the exact physical overdensity of each absorber at
$R$. The physical overdensity of an absorber at $R$ is $\delta\propto
\left<\Delta(R)\right>$.  The detailed values used for the involved parameters
come empirically from the comparison of our simulation results with the
observations below.  While this should be sufficient for the purpose of this
paper, those constraints may be improved by future high-resolution spectra of
QSOs and the PEs to be revealed and the comparison with cosmological
simulations of PEs with including detailed 3D radiative transfer processes.
\label{ft:DeltaR} }.
As shown in panels (a) and (b), the histogram is systematically and substantially
higher than the blue line but consistent with the green line (no LOSPE), which
suggests that the sample variance is highly unlikely to be responsible for the
excess of absorption in the QSO near zone and the density in the QSO near
zones is enhanced. Indeed, the expected $DA$ can be consistent with the
observations if the density is enhanced in the QSO near zone (as
shown by the red line and points in panels a and b).  Note also that the error to
the mean of $DA$ for a sample of 500 spectra is sufficiently small, which
suggests that the physical properties of QSOs can be extracted from the PE for
samples with 500 spectra or more.

The PE analysis is sensitive to the accuracy in the systemic redshift
estimation of each QSO/fgQSO being studied if the size of its proximity region
is smaller than or comparable to the scale corresponding to the systemic
redshift error. However, the systemic redshift estimated from the UV emission
lines of a QSO is generally systematically smaller than the systemic redshift
of its host galaxy by more than a few hundred kilometer per second.
\citet{Richards02} find that CIV gives a redshift smaller than Mg II by
824~$\kms$ with a scatter of $511\kms$ from a subset of SDSS QSO spectra.
\citet{Nestor08} find that the [OIII], which should better represent the
systemic redshift, gives a systemic redshift larger than Mg II by $102~\kms$.
According to these results, the errors in the systemic redshift estimates
of QSOs can be divided into two parts, i.e., the mean systematic offset
and the scatter around the mean offset, which are denoted as the ``systematic error''
and the ``random error'' below, respectively. According to \citet{Richards02}
and \citet{Nestor08}, we adopt a systematic error  of $926\kms$ in the estimates of
the fgQSO systemic redshifts and a random error with dispersion $511\kms$ (as the redshift of
the majority of the QSOs in KT08 is estimated by CIV), unless otherwise
specified. Note that the systematic error adopted in KT08 is smaller, $\sim753\kms$,
and the dispersion of the random errors is $676\kms$, and we also try those
errors in our simulations below (i.e., in Figure~\ref{fig:f5}, where the systemic
redshift of each mock QSO is set to be a value of $z_{\rm fg}=2$ plus a Gaussian
distributed random error). In order to
compare with the observational results from KT08, an additional correction in
the systematic error of $\left<\delta z_{\rm offset}\right>=173\kms$ needs to
be taken into account, i.e., the systemic redshift of each QSO host in the mock samples is set to
$z_{\rm fg}+ \delta z_{\rm offset}$, where $z_{\rm fg}$ is set to $2$ and
the random error in $\delta z_{\rm offset}$ is also assumed to be Gaussian distributed
(e.g., in Figures~\ref{fig:f2}, \ref{fig:f3}d, \ref{fig:f4}, \ref{fig:f6}, \ref{fig:f7} below). For
comparison, a zero error of $\delta z_{\rm offset}$ is assumed in Figure~\ref{fig:f3}a-c below.
Generally, the larger the systematic error in the systemic redshift estimates, the less the enhancement
of the density in the fgQSO near zones that is required to reproduce
the observed LOSPE (see Figure~\ref{fig:f2}). 

\subsubsection{The TPE of type 1 fgQSOs}\label{sec:tpe1}

\begin{figure*}
\includegraphics[width=6.5in]{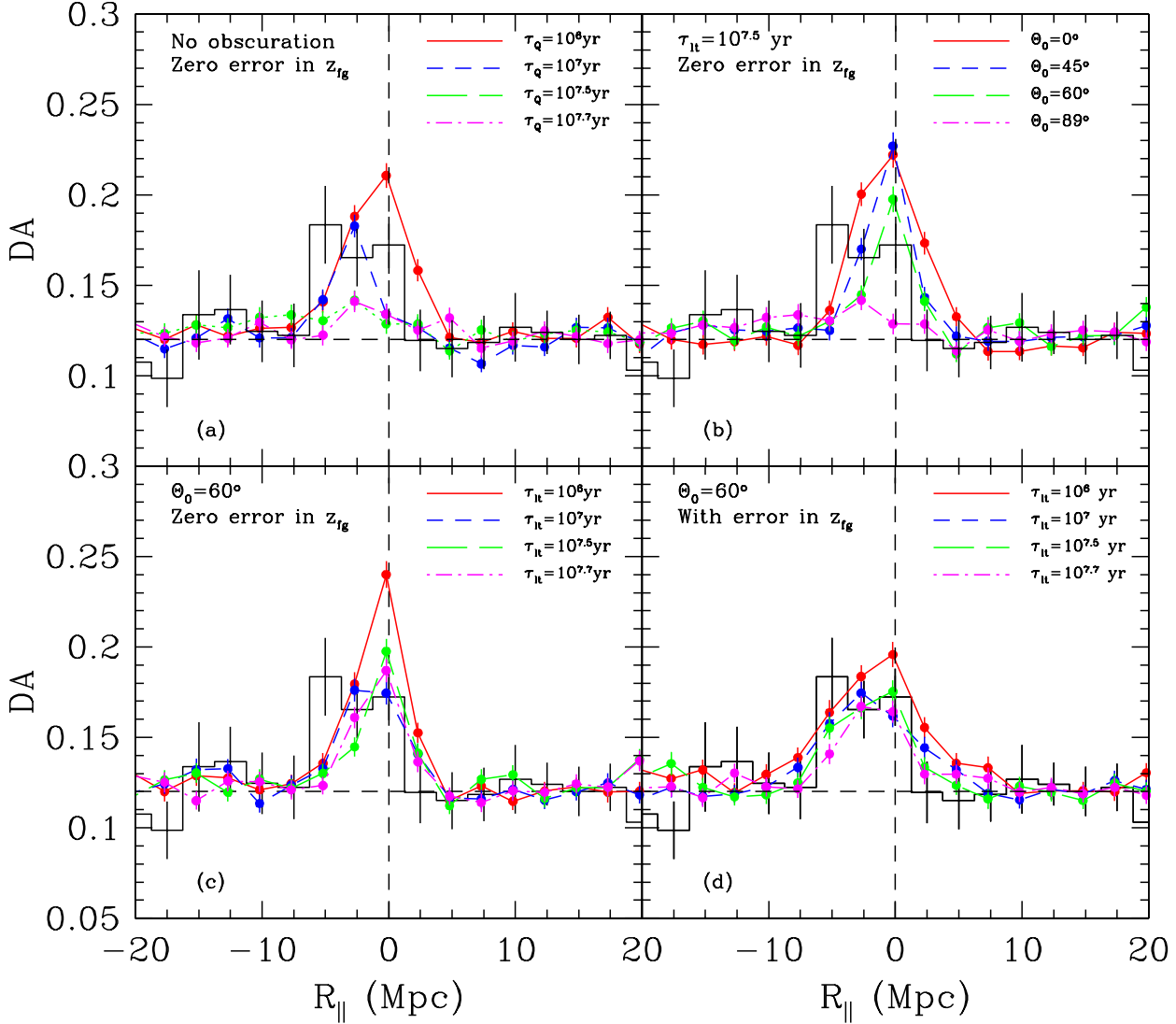}
\caption{
The transverse proximity effect for type 1 foreground QSOs at redshift $\sim
2$. The histogram shown in each panel represents the average flux decrement
$DA$ obtained by KT08 for a sample of background QSOs as a function of
proper distances along the sightline, with the origin corresponding to the
redshift of the paired fgQSOs; and other lines and symbols represent our
simulation results. In panel (a), the simulation results of $DA$ is obtained
under the assumption that there is no obscuration to the fgQSO UV radiation and
no error in the estimates of the systemic redshifts of fgQSOs.  Different
colors represent different QSO ages, as labeled in the panel.  Panel (b) shows
the simulation results obtained by assuming that the QSO lifetime is
$10^{7.5}$yr, and the QSO age is randomly chosen over a range from 0 to each
assumed QSO lifetime.  Different colors represent different half-opening angles
of the torus associated with each fgQSO $\Theta_0$.  Panel (c) shows the
simulation results obtained with $\Theta_0 =60\degr  $ and different values of
QSO lifetime.  All the simulations in panels (a)--(c) assume a zero error in
the estimates of the fgQSO systemic redshift.  Panel (d) show the simulated
$DA$ obtained by the same assumptions as those in panel (c), except for
accounting for both the systematic and random errors in the estimates of the fgQSO systemic redshift. All the
circle points shown in the figure are centered on the same bins of $R_{\parallel}$ as the
observational data but have been slightly offset to the left for graphical
clarity (similarly in Figures~\ref{fig:f5} and \ref{fig:f6} below).
}
\label{fig:f3}
\end{figure*}

Assuming that the effective density enhancement in the near zones of the fgQSOs
is the same as the one required to reproduce the observations on the LOSPE by
KT08\footnote{We assume that the density enhancement in the QSO near zone is
isotropic. We do not consider an anisotropic distribution of the density
enhancement in this paper.}, we generate mock samples with 500 synthetic
Ly$\alpha$ forest spectra and obtain the TPE effect from these samples.
Figure~\ref{fig:f3} shows the expected TPE obtained from the mock samples with
different settings of the QSO age $\tauq$, the opening angle of the torus
$\Theta_0$, and the error in the estimation of the fgQSO systemic redshift.
Figure~\ref{fig:f3}a shows that for a small QSO age $\tauq$, the excess of $DA$
is significant near the fgQSOs because of the significant density enhancement
in the immediate vicinity of the fgQSOs. This excess of $DA$ decreases with
increasing $\tauq$ because the significance of the suppression of the
absorption by the UV photons from the fgQSOs increases with increasing $\tauq$.
In principle, the dependence of the TPE on the QSO age $\tauq$, as shown in
Figure~\ref{fig:f3}a, suggests that the TPE can be used to constrain the QSO
lifetime once the density enhancement in the fgQSO near zones is
determined by the LOSPE. The differences among large $\tauq$ cases shown in the
panel are small because the time during which photons cross the proximity regions are
smaller than or at most comparable to the QSO age in these cases.  However,
none of the simulated $DA$ (even the one with the shortest $\tauq$) in
Figure~\ref{fig:f3}a can match the observational results on the TPE obtained by
KT08. 

Figure~\ref{fig:f3}b shows the dependence of the TPE on the half opening angle
$\Theta_0$ of the tori associated with the fgQSOs. For the extreme case of
$\Theta_0=0\degr  $ (without the PE due to the fgQSO UV radiation), the excess
of the $DA$ due to the density enhancement near fgQSOs is the most significant
(red line and points). With increasing $\Theta_0$, the excess of $DA$ becomes
less and less significant because the region that can be affected by the UV
photons escaping out from the central engine becomes larger as shown by the
color lines and points.  For $\Theta_0=89\degr  $ (almost without obscuration
to the fgQSO photons; magenta line and points), the expected excess of $DA$ due
to the density enhancement is balanced significantly by the proximity effect
due to the fgQSO UV radiation.

Figure~\ref{fig:f3}c shows the dependence of the TPE on the lifetime of fgQSOs,
in which the half opening angle of the torus is fixed to $\Theta_0=60\degr  $.
This value of $\Theta_0$ is roughly in the range determined by observations
\citep[e.g.,][]{TU06,Gilli07,Treister10}. In this panel, the value of the QSO
age $\tauq$ is chosen randomly over a range from 0 to the QSO lifetime
$\tau_{\rm lt}$, instead of being a constant as labeled for each line in panel
(a).  Because of the obscuration in the transverse direction to the UV
radiation from the fgQSOs, there is significant $DA$ excess even for the case
of $\tau_{\rm lt}=10^{7.7}$~yr. Compared to the case without obscuration shown
in Figure~\ref{fig:f3}a, the dependence of the $DA$ excess on $\tau_{\rm lt}$
becomes less obvious if the obscuration to the UV radiation from the fgQSOs is
significant.

Note that one characteristic timescale for the luminosity evolution of a QSO is
the Salpeter timescale $\tau_{\rm sp}$, which is $ \sim 10^{7.7}$~yr if the
mass-to-energy conversion efficiency of its nuclear activity is $\sim 0.1$ and
the Eddington ratio is $1$. If the age of a fgQSO is larger than the Salpeter
timescale, the proximity effect may be substantially less significant for
regions faraway from the fgQSOs with $R(1-\cos\theta_*)\ga c\tau_{\rm sp}$,
considering the luminosity evolution of the fgQSOs may be significant.

Figure~\ref{fig:f3}d shows the expected TPE obtained from the mock samples
after taking into account the errors in the systemic redshift of the fgQSOs. As
in panel (c), the opening angle of the tori associated with the fgQSOs is fixed
to $60\degr  $ and the age of fgQSOs $\tauq$ is randomly chosen over a range of
$(0,\tau_{\rm lt})$. As seen from the panel, the observations by KT08 can be
well reproduced if the lifetime of fgQSOs is $\sim 10^{7.5}$~yr (green line
and points). Our calculations also show that the expected $DA$ for cases with
$\tau_{\rm lt}\la 10^6$~yr cannot match the observations for any given
$\Theta_0$, which is different from the suggestion that the fgQSOs have had
their current UV luminosities for less than approximately a million years made
in KT08.  Our results suggest that the lifetime of the fgQSOs is consistent
with being a few times $10^7$~yr and a substantially smaller lifetime ($\la
10^6$~yr) is not required by the PE, and it is also consistent with those
constraints obtained by the detections of the TPE through He~{\small II}
Ly$\alpha$ forest or metal absorption lines
\citep[e.g.,][]{Jakobsen03,WW06,Worseck07,GSP08}

Note here that the two components of some QSO pairs in KT08 are at similar
redshift and
therefore the TPE of those fgQSOs may be affected by the bgQSOs.  The UV light
from the bgQSO near a fgQSO may contribute some to the ionization of the near
zone of the fgQSO, but this effect should be more significant at the backside
than the front side of the fgQSO and lead to a more significant decrease in the
$DA$ at $R_{\parallel}\sim -3$ -- $-5$~Mpc.  This effect cannot explain the
significant excess of $DA$ at $R_{\parallel}\sim -3$ -- $-5$~Mpc but rather
require a more significantly enhanced density in the fgQSO near zone. In
addition, if the density is also enhanced in the near zone of the bgQSOs, a
more significant excess of the absorption at $R_{\parallel}\sim -5$~Mpc can be
expected, which might bring the observed $DA$ at $R_{\parallel}=-5$~Mpc in
better consistency and strengthen the above conclusions.

\begin{figure}
\includegraphics[width=84mm]{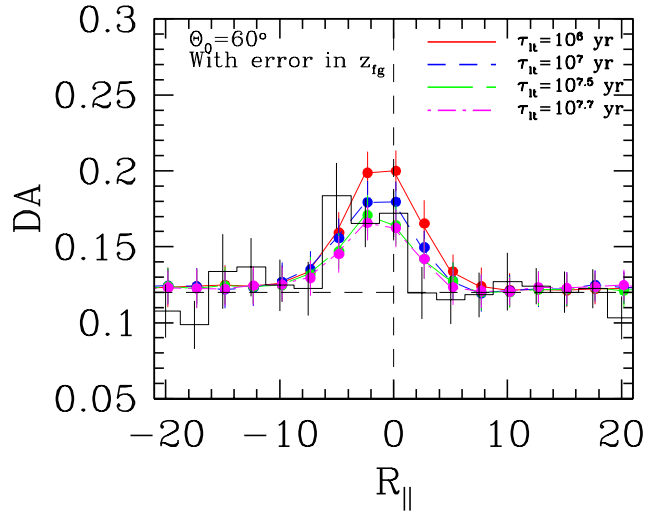}
\caption{The transverse proximity effect
of fgQSOs expected from 100 realizations of mock samples each with
130 synthetic Ly$\alpha$ forest spectra. Legends are similar to that
for Figure~\ref{fig:f3}d, except the errorbar to each bin is the
standard deviation of the sample.
}
\label{fig:f4}
\end{figure}

In order to check the significance of the sample variance on the TPE obtained by
KT08, we again generate 100 realizations of mock samples, each with 130
synthetic Ly$\alpha$ spectra. Figure~\ref{fig:f4} shows the expected TPE for
those mock samples with similar settings to that shown in Figure~\ref{fig:f3}d.
According to this Figure, the KT08 results are still consistent with that the
QSO lifetime is as long as a few $10^7$~yr and the torus half opening angle
$\Theta_0\sim 60\arcdeg$. For the cases of $\tau_{\rm lt} \la 10^6$~yr, the 
probability that the inconsistency between the KT08 TPE results and the
expected TPE are simply due to sample variance is low.

There are still some uncertainties in the current estimation of UVB which may
affect the simulation results. In our calculations, we set the UVB as
$\Gamma_{\rm UVB} =10^{-12}{\rm s}^{-1}$, which might be somewhat large as the
latest estimation of \citet{FGetal08a} is only $0.5\times
10^{-12}{\rm s}^{-1}$.  If the UVB is set to be this smaller value, a more
significant LOSPE would be expected, and the effective density enhancement
factor $\left<\Delta(R)\right>$ should be slightly larger compared to that in
Section~\ref{sec:low} in order to fit the LOSPE estimated by KT08.  Therefore,
the QSO lifetime $\tau_{\rm lt}$ and/or the torus opening angle $\Theta_0$ are
required to be even larger than that given in Section~\ref{sec:low} in order to
reproduce the TPE. But if the UVB is unreasonably much larger than
$10^{-12}{\rm s}^{-1}$, the excess of absorption at the transverse direction
detected by KT08 cannot be explained by simply changing $\tau_{\rm lt}$,
$\Theta_0$ and $\left<\Delta(R)\right>$.

We caution here that other uncertainties could also affect the results
presented here quantitatively. First, the $DA$ estimated from observations may
be affected by the continuum fitting. However, the error in the continuum fitting is
typically on the percentage level depending on the signal-to-noise ratio of the
QSO spectrum \citep{Kim07}, which is not likely to change the observational
results by KT08 on PE qualitatively. Second, the combined sample in KT08 are
obtained from several different instruments and is highly heterogeneous. The
ignoring of the detailed exact redshift and luminosity distributions of the QSO
pairs in our simulations is sufficient for the demonstration purpose in this
paper, but it may introduce some uncertainties to the resulted PE. Future works
on extracting QSO properties from the PEs should consider the uncertainties.
In addition, the luminosity of QSOs may evolve or fluctuate on timescales
$10^6$~yr, which may lead to more significant excess of absorption in the
transverse directions \citep{Adelberger04}. However, such a luminosity
variation of QSOs is not required according to our simulations.

\begin{figure}
\includegraphics[width=84mm]{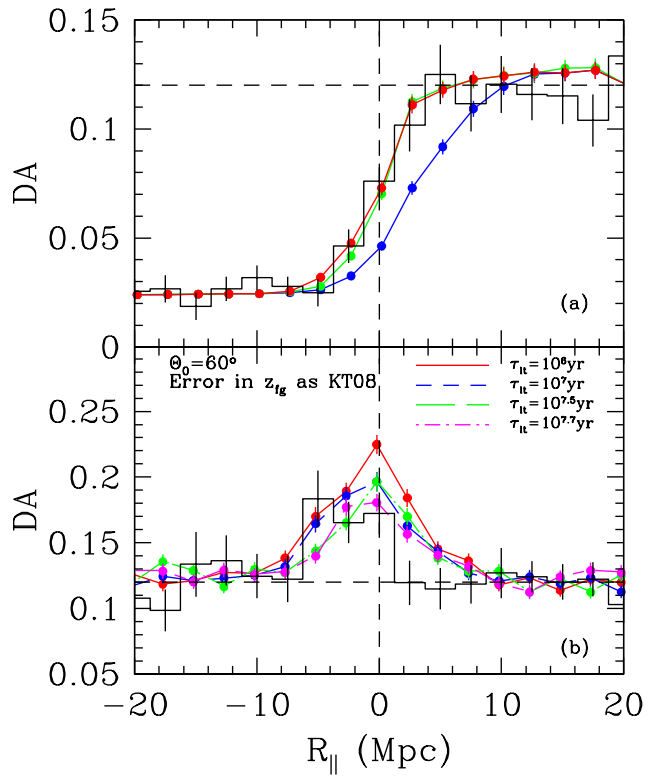}
\caption{The line of sight proximity effect and the transverse proximity effect
of fgQSOs that both the systematic and random errors in their systemic redshift estimation used are the
same as that in KT08. Legends of panel (a) are similar to those of
Figure~\ref{fig:f3} and legends of panel (b) are similar to those of
Figure~\ref{fig:f3}d. }
\label{fig:f5}
\end{figure}

Figure~\ref{fig:f5} shows the expected $DA$ obtained from the mock samples by
assuming that both the systematic and random errors in the estimates of the fgQSO systemic redshifts
through CIV are the same as those adopted in KT08. Compared to the density
enhancement used in Figure~\ref{fig:f2}, here a larger value of it is used so
that the LOSPE of fgQSOs obtained by KT08 can also be re-produced well (see the
red line and points in Figure~\ref{fig:f5}a).  With the density enhancement
required by the LOSPE, the $DA$ distribution near the fgQSOs indicated by the
bgQSO spectra (i.e., the TPE) is also calculated as shown in
Figure~\ref{fig:f5}b. We find that the observational asymmetric distribution of
the excess in $DA$ near $R_{\parallel}\sim 0$, i.e., the lack of excess in $DA$
at $R_{\parallel}\sim 2.5$~Mpc and the significant excess of $DA$ at
$R_{\parallel}\sim -5$~Mpc, cannot be simultaneously re-produced for any given
$\tau_{\rm lt}$ and $\Theta_0$, in contrast to the results in
Figures~\ref{fig:f3}d and \ref{fig:f4}.

\subsubsection{The TPE of type 2 fgQSOs}\label{sec:tpet2lz}

\begin{figure}
\includegraphics[width=84mm]{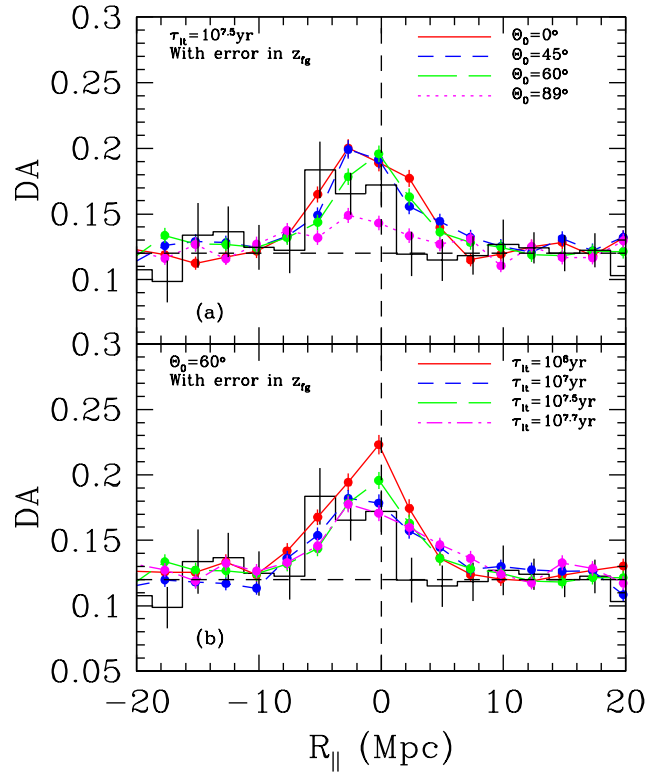}
\caption{ The transverse proximity effect for type 2 fgQSOs with the same
intrinsic luminosity $L_{\nu_0}$ at redshift $\sim 2$ as those in
Figure~\ref{fig:f3}b and d.  Legends for panel (a) and (b) are similar to those
of Figure~\ref{fig:f3}b and d, respectively.  As a reference, the excess
absorption near type 1 fgQSOs measured by KT08 is also shown by the
histogram here.
}
\label{fig:f6}
\end{figure}

\begin{figure}
\includegraphics[width=84mm]{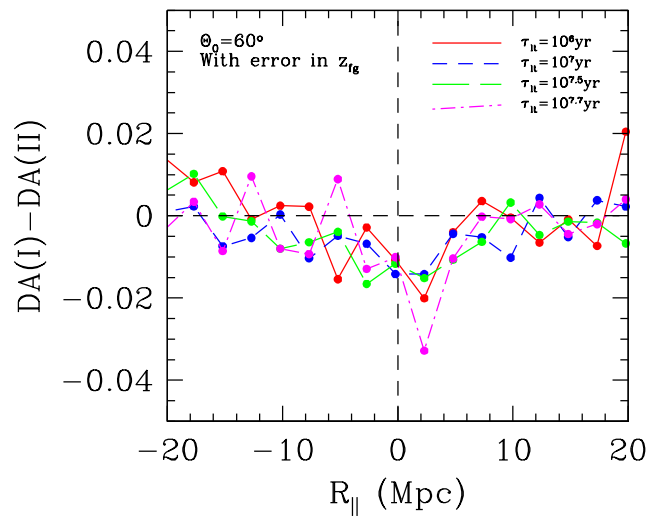}
\caption{The difference of the transverse proximity effect between type 1 fgQSOs
and type 2 fgQSOs shown in Figure~\ref{fig:f3}d and Figure~\ref{fig:f6}b.
The flux decrement in the
proximity region of type 1 fgQSOs and type 2 fgQSOs are represented by $DA$(I)
and $DA$(II), respectively. Legends are similar to those in Figure~\ref{fig:f3}d.
}
\label{fig:f7}
\end{figure}

Assuming that the effect of the density enhancement near type 2 fgQSOs is the
same as that of type 1 fgQSOs with similar intrinsic $L_{\nu_0}$ constrained by
the LOSPE (see Figure~\ref{fig:f2}), we generate synthetic Ly$\alpha$ forest
spectra of type 1 bgQSOs whose light passed by the proximity region of type 2
fgQSOs to study the TPE of type 2 fgQSOs. Figure~\ref{fig:f6} shows our
simulation results on their $DA$ distribution near the fgQSOs.
Figure~\ref{fig:f6}a shows the dependence of the TPE on the half opening angle
of the torus $\Theta_0$ for the mock samples of type 2 fgQSOs, given the QSO
lifetime. As seen from the panel, generally the larger the $\Theta_0$, the less
the excess of the absorption near $R_{\parallel}\sim 0$~Mpc.
Figure~\ref{fig:f6}b shows the dependence of the TPE on the QSO lifetime, given
the half opening angle $\Theta_0=60\degr  $.  As seen from the panel, the
differences in $DA$ are relatively small among the cases with large $\tau_{\rm
lt}$ ($\sim 10^7-10^{7.7}$~yr). If $\tau_{\rm lt}\sim 10^6$~yr, the absorption
at $R_{\parallel}\sim 0$~Mpc is larger than those cases with large $\tau_{\rm
lt}$ due to the time-delay effect.  Generally the absorption in the region
closer to the observer ($R_{\parallel}>0$) should be relatively large compared
to that for type 1 fgQSOs (see Figure~\ref{fig:f3}d), because the near side is
more likely to be obscured from the fgQSOs. That effect, though weak, is shown
in Figure~\ref{fig:f7} (see some relative deep dips of the curves at
$R_{\parallel}\sim2.5$~Mpc).  This contrast between the absorption $DA$ of type
2 fgQSOs and that of type 1 fgQSOs appears more significant for higher
luminosity mock samples, as will be seen in Section~\ref{sec:tpe2hz} and
Figure~\ref{fig:f11} below.  

Note that in this paper, the UV radiation is assumed to be intrinsically
isotropic, after removing the anisotropic effect due to the torus.  However, if
the UV radiation is intrinsically anisotropic, the TPE of type 2 fgQSOs may be
different from those expected from the above calculations.  For example, if the
UV (and X-ray) radiation is relatively stronger along the direction of the
torus axis than in the direction along the torus plane, the expected $DA$ excess
for type 2 fgQSOs should be smaller at $R_{\parallel}\sim 0$~Mpc than those calculated
above.

\subsection{High-luminosity samples at $z\sim 4$}\label{sec:high}

\subsubsection{The LOSPE of type 1 fgQSOs} \label{sec:lospehz}

\begin{figure}
\includegraphics[width=84mm]{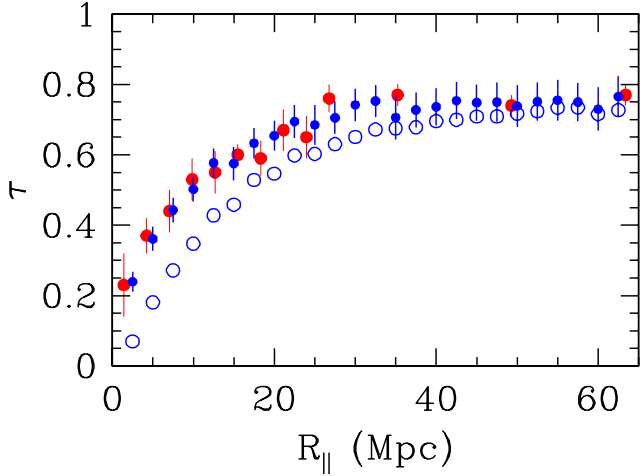}
\caption{The line of sight proximity effect for type 1 QSOs.  The red points
show the optical depth distribution in the QSO proximity region obtained by
\citet{Guimaraes07} for an observational sample of QSOs with a mean Lyman limit
luminosity of $\sim 8\times 10^{31} \ergshz$ at redshift $\sim 4$. The blue
open circles represent the optical depth distribution obtained from our
simulated mock sample by assuming that the density in the QSO proximity regions
is the same as that of the cosmic average. The blue solid circles with
errorbars represent the simulated results by assuming that the density
distribution in the QSO proximity region is enhanced by a factor of
$\Delta(R)$ (see details in Section~\ref{sec:lospehz}).  }
\label{fig:f8}
\end{figure}

Figure~\ref{fig:f8} shows the optical depth distribution near QSOs/fgQSOs along
their LOSs with the origin corresponding to their redshift (similar to
Figure~\ref{fig:f2}).  The red points with errorbars show the results obtained
by \citet{Guimaraes07} for an observational sample of $\sim 50$ high-redshift
QSOs ($z\sim 4$) with mean Lyman limit luminosity of $L_{\nu_0}\sim 8\times
10^{31} \ergshz$. The blue open circles represent the median optical depth
obtained from our mock sample with 500 spectra affected by fgQSOs by assuming
that the density distribution near the fgQSOs is the same as that of the cosmic
average, which are clearly offset from the observations.  Note that the errors
in the estimates of the fgQSO systemic redshift are also considered in our
simulation here by adopting the method same as that used in
\citet{Guimaraes07}, i.e., the error is randomly chosen over the range of
$0-1500\kms$. The decrease of the simulated optical depth with decreasing
distance to the QSOs appears more significant than the observation results, as
the effect of the density enhancement in the proximity regions of these QSOs is
ignored. Assuming that the density in the QSO proximity regions is enhanced by
a factor of $\left<\Delta(R)\right>\sim 1+ C_0(R/C_R)^{-p} \exp[-(R/C_R)^q]$,
the observations can be reproduced by simulations (blue solid circles) if
$C_0=0.9$, $C_R=11$~Mpc, $p=0.7$, and $q=0.5$.\footnote{With the adopted
parameters here, the effective density enhancement is almost the same as that
required for the low-luminosity sample of KT08 at $R<3$~Mpc but declines slower
at $R>3$~Mpc (see Section~\ref{sec:low}). } For this case, we set the number of
spectra in each realization to be 50 and we simulate 100 realizations. Thus the
mean median optical depth and its standard deviation can be
estimated from these realizations. As seen from Figure~\ref{fig:f8}, the
observations can be well matched by the simulation results (blue solid points).

\citet{Guimaraes07} also suggest that the density enhancement in the near zones
of QSOs is required in order to explain their measurements on the LOSPE. Compared
with their estimates on the density enhancement in the near zones of those high
luminosity QSOs, our estimates are smaller by a factor of $1.5-5$ at a distance 
of $R_{||}\la 10$~Mpc-$1$~Mpc but similar at a distance of $R_{||}\ga 15$~Mpc.
The main reason for this difference is that the Poisson variance of absorption
lines in different sight lines to 
the near zones of QSOs, which is included in the Monte-Carlo simulations here,
leads to an averaged absorption that is larger than a simple estimation
obtained without
considering of the Poisson variance \citep[see][]{Guimaraes07}. Therefore,
the density enhancement required here is less significant compared with that obtained by
\citet{Guimaraes07} (see also discussions on the Poisson variance in \citealt{Dall08}).

\subsubsection{The TPE of type 1 fgQSOs}\label{sec:hztpe1}

\begin{figure*}
\includegraphics[width=6.5in]{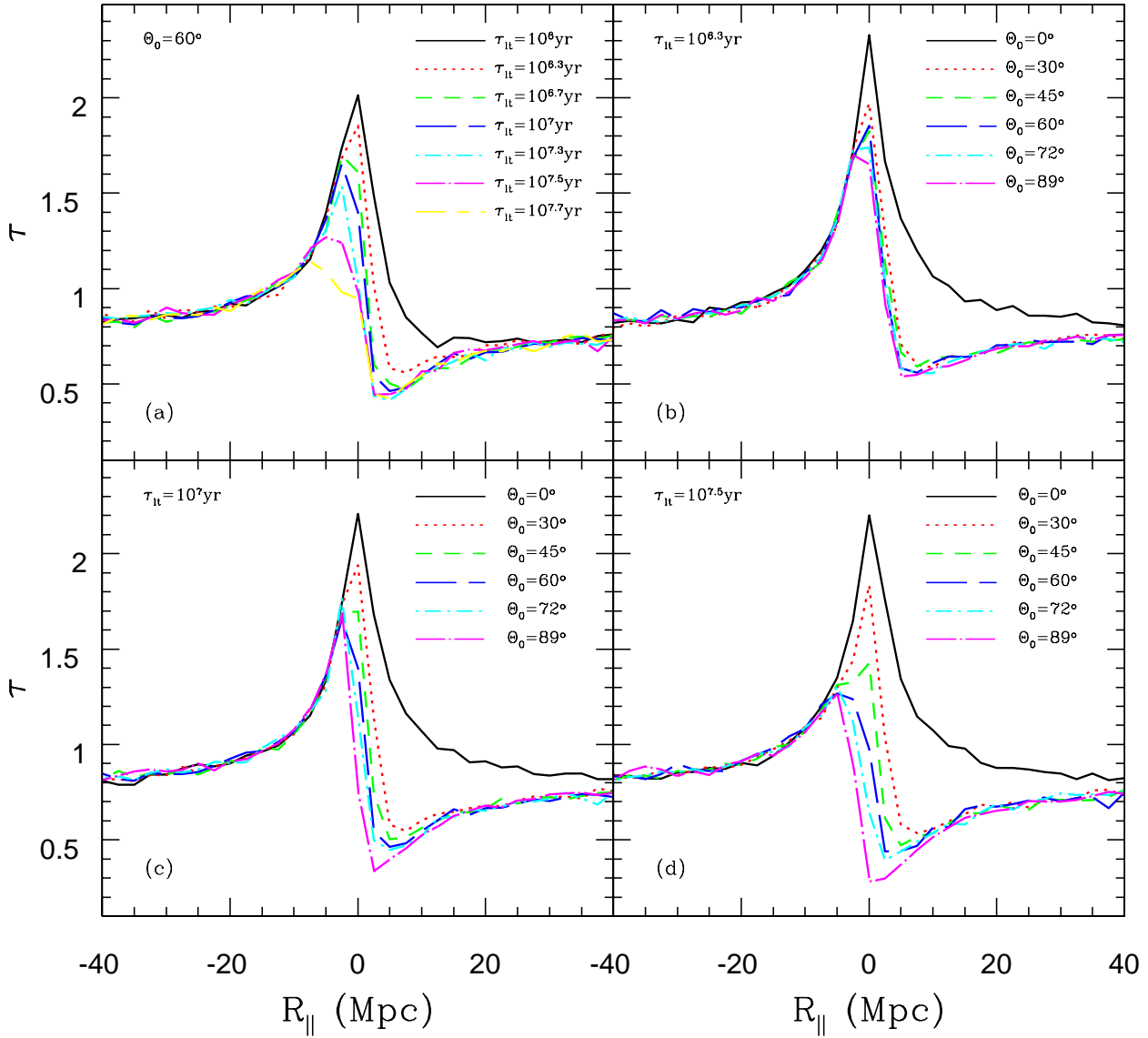}
\caption{The expected transverse proximity effect of type 1 fgQSOs at redshift
$\sim 4$ with Lyman limit luminosity of $L_{\nu_0}\sim 8\times 10^{31}\ergshz$.
Panel (a) shows the expected optical depth distribution in the proximity region
of fgQSOs, in which the half opening angle of each fgQSO is assumed to be
$60\degr  $ and different curves represent the results obtained for different
sets of the fgQSO lifetime $\tau_{\rm lt}$. The other three panels show the
dependence of the optical depth on the half opening angle of the torus, given
the fgQSO lifetime $\tau_{\rm lt}=10^{6.3},10^7,10^{7.5}$, respectively.  }
\label{fig:f9}
\end{figure*}

Although so far there is no observational measurement on the TPE of QSOs (with
luminosity $\sim 8\times 10^{31}\ergshz$) similar to the sample in
\citet{Guimaraes07}, we illustrate here the effects of different QSO properties
on the TPE and demonstrate that both $\tau_{\rm lt}$ and $\Theta_0$ can be
simultaneously constrained by the TPE if the density enhancement in the fgQSO
proximity region has been constrained by the LOSPE. We first assume that the
effect of the density enhancement in the proximity regions of those QSOs is the
same as that indicated by the LOSPE obtained by \citet{Guimaraes07} (see Figure
\ref{fig:f8}). Under this assumption, we synthesize a large number of
Ly$\alpha$ forest spectra of bgQSOs with different choices of the QSO lifetime
and the half opening angle of the associated torus (see parameter settings in
Section~\ref{sec:Geom}). We then extract the median optical depth distribution
in the fgQSO proximity regions from these spectra and show the results in
Figure~\ref{fig:f9}. As seen from Figure~\ref{fig:f9}a, given the half opening
angle of the torus associated with the fgQSO, e.g., $\Theta_0=60\degr  $,
generally the optical depth in the proximity region to the QSO (e.g.,
$|R_{\parallel}|\la 10$~Mpc) decrease with increasing the QSO lifetime; and the
decrease is significant initially at the frontside to the QSO, and becomes
significant at its backside (e.g., $R_{\parallel}\sim-4$~Mpc) when the QSO
lifetime $\tau_{\rm lt}$ is long enough so that the region can be reached by
the radiation from the fgQSO.  That tendency is manifest especially for large
$\Theta_0$, as (1) for the case with short QSO lifetime (e.g., $\tau_{\rm
lt}\sim 10^{6.3}$~yr; see Figure~\ref{fig:f9}b), the insignificant decrease of
the optical depth is not sensitive to different choices of $\Theta_0$; and (2)
as shown in Figure~\ref{fig:f9}c and d, for large $\tau_{\rm lt}$, the decrease
of the optical depth at $R_{\parallel}\la 0$~Mpc increases with increasing
large $\Theta_0$.  As shown in the figure, the QSO lifetime and the half
opening angle of the torus affect the optical depth curves in different ways,
so that they can be simultaneously constrained by the LOSPE and TPE of QSOs if
$\tau_{\rm lt}$ and $\Theta_0$ are not too small (e.g., $\tau_{\rm lt} \ga
10^6$~yr and $\Theta_0\ga 30\degr  $). Note here that the standard deviation,
on the order similar to that shown in Figure~\ref{fig:f9}, is small enough so
that the constraints on the QSO lifetime and the torus half opening angle may
be accurately extracted through the PEs of a sample with a few hundreds of high
luminosity QSOs.

Considering the possibility of inaccuracy in the errors of the QSO systemic
redshifts used above, we also test that our results obtained above remain the
same qualitatively, if the systemic redshift errors is set to be the same as
that for the low-luminosity low-redshift sample in Section~\ref{sec:tpe1}.  The
overall shape of the optical depth distribution around $R_{\parallel}\sim0$~Mpc
does not change significantly, though it may shift slightly toward the negative
$R_{\parallel}$ direction, as discussed in Section~\ref{sec:tpe1}. 

\subsubsection{The TPE of type 2 fgQSOs} \label{sec:tpe2hz}

\begin{figure}
\includegraphics[width=84mm]{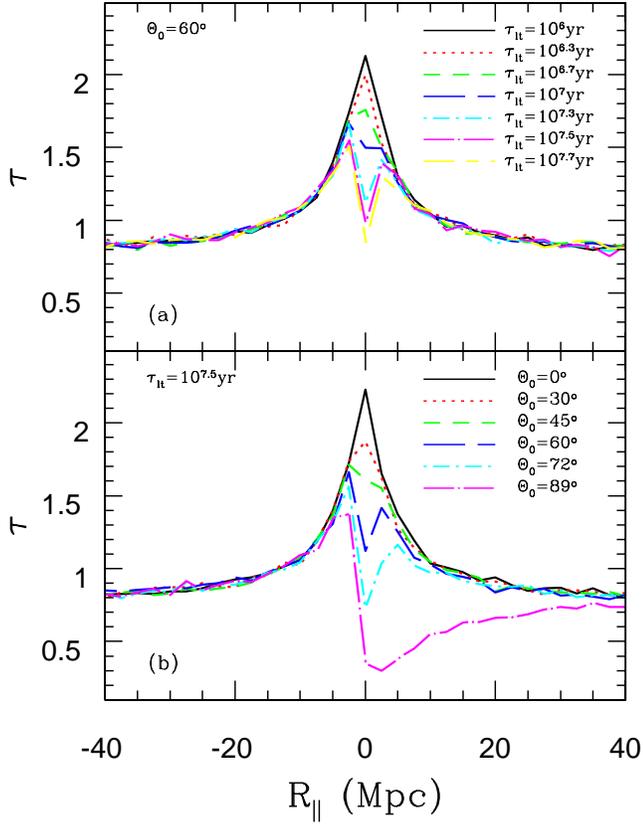}
\caption{The expected transverse proximity effect of type 2 QSOs at redshift
$\sim 4$.  Legends for panels (a) and (b) are the same as those for
Figure~\ref{fig:f9}a and d, respectively. See details in Section~\ref{sec:tpe2hz}. }
\label{fig:f10}
\end{figure}

Assuming that the effect of the density enhancement near type 2 fgQSOs is the
same as that of type 1 fgQSOs with similar $L_{\nu_0}\sim 8\times 10^{31}
\ergshz$ at $z\sim 4$, we show our simulated TPE of these type 2 fgQSOs in
Figure~\ref{fig:f10}. As seen from Figure~\ref{fig:f10}a, given the half opening
angle of the torus (e.g., $\Theta_0=60\degr  $), the larger the QSO lifetime
$\tau_{\rm lt}$, the smaller the optical depth at $R_{\parallel}\sim 0$~Mpc
because the UV photons from the fgQSOs are more likely to propagate to more
distance regions in the transverse direction. Given the QSO lifetime, the
larger the half opening angle of the torus, the smaller the optical depth at
$R_{\parallel} \ga 0$~Mpc (Figure~\ref{fig:f10}b).

\begin{figure}
\includegraphics[width=84mm]{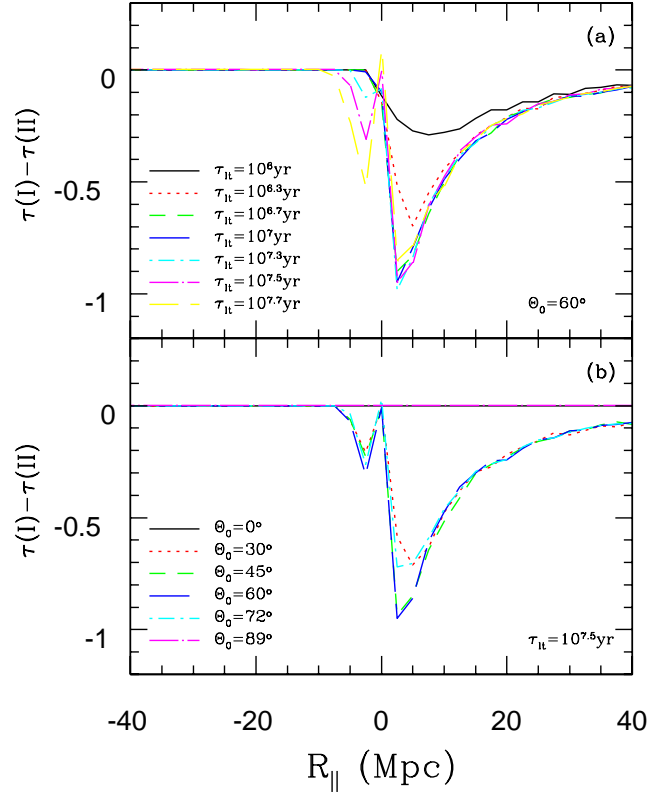}
\caption{The expected difference in the transverse proximity effect of type 1
fgQSOs at redshift $\sim 4$ from that of type 2 fgQSOs. The optical depth
distribution near type 1 fgQSOs is denoted by $\tau$(I), and  that near
type
2 fgQSOs denoted by $\tau$(II).  Legends are the same as those for
Figure~\ref{fig:f10}. See details in Section~\ref{sec:tpe2hz}. }
\label{fig:f11}
\end{figure}

Figure~\ref{fig:f11} shows the difference between the optical depth
distribution near type 2 fgQSOs and that near type 1 fgQSOs, assuming that
other parameters of these two types of fgQSOs (e.g., $\tau_{\rm lt}$ and
$\Theta_0$) are the same. In contrast to the cases considered for fgQSOs with
$L_{\nu_0}\sim5\times 10^{30}\ergshz$ at redshift $\sim 2$, the difference here
is obvious at the near side of the fgQSOs and for some cases in the far side
because the fgQSOs in the mock samples here are much more luminous than that in
Section~\ref{sec:tpet2lz}. Given the half opening angle of the torus,
$\Theta_0=60\degr  $, the optical depth difference at $R_{\parallel} \sim
0-15$~Mpc increases with increasing $\tau_{\rm lt}$ but saturates when
$\tau_{\rm lt}\ga$ a few times $10^7$~yr; while the difference at the far side
$R_{\parallel}<0$ only becomes visible when $\tau_{\rm lt}\ga10^7$~yr.  Given a
QSO lifetime, the optical depth difference is close to 0 for either small half
opening angle of torus $\Theta_0\rightarrow 0\degr  $ or large
$\Theta_0\rightarrow 90\degr  $; however, this difference is the largest for a
medium $\Theta_0\sim 60\degr  $.  That behavior of the optical depth difference
is the combined effects due to the geometrical nature of the torus and the time
delay of photons that propagated to the backside of fgQSOs.  The sharp contrast
of the TPE due to type 1 fgQSOs from that due to type 2 fgQSOs should be useful
to probe the properties of QSOs through the PE.

The simulated results illustrated above for the fgQSOs similar to that in the
sample of \citet{Guimaraes07} again suggest that the QSO properties, such as
the density enhancement in its proximity region, the lifetime and the opening
angle of the associated torus, can be constrained simultaneously by the LOSPE
and TPE of type 1 (and/or type 2) fgQSOs. As the fgQSOs in these samples are
much more luminous, their TPE is obvious, without being smeared out by the
errors in the systemic redshift estimates of the fgQSOs.

\section{Conclusions and discussions}\label{sec:conclusion}

In this paper, we have investigated both the LOSPE and the TPE due to type 1
QSOs and the TPE due to type 2 QSOs. To do so, we adopted the Monte-Carlo
method to generate a large number of simulated Ly$\alpha$ forest spectra of
type 1 bgQSOs that may be affected by (fg)QSOs, according to the statistical
distributions of the HI column density, the Doppler parameter of the Ly$\alpha$
absorption lines, and the number evolution of the absorbers associated with
these absorption lines. We extracted the distribution of the flux decrement
($DA$) or the optical depth near the fgQSOs from these spectra.  We illustrated
that the effects of the fgQSOs on the bgQSO spectra depend on the fgQSO
lifetime, the anisotropy in the fgQSO UV radiation, and the density enhancement
in the fgQSO proximity region, given the strength of the UVB.  According to
these calculations, the PE of QSOs may be divided into the ionization proximity
effect (IPE) and the density proximity effect (DPE).  The absorption near a QSO
may be enhanced due to the DPE but decrease due to the IPE. Depending on
detailed properties of QSOs, the combination of DPE and IPE can result in
either an excess or a decrease of absorption in the near zone of a fgQSO.

The LOSPE of a QSO with known luminosity is affected by the density enhancement
in the QSO near zone but irrelevant to the QSO lifetime and the anisotropy
in the QSO UV radiation, which can thus provide constraint on the effective
density enhancement in the near zone of the QSO. Based on the measurements of
LOSPE for a low-redshift low-luminosity QSO sample (i.e., $z\sim 2$ and
$L_{\nu_0} \sim 5\times 10^{31}\ergshz$; see KT08) or a high-redshift high
luminosity QSO sample (i.e., $z\sim 4$ and $L_{\nu_0}\sim 8\times
10^{31}\ergshz$; see \citealt{Guimaraes07}), we obtained the effective density
enhancement in the near zones of the QSOs in these two samples by matching the
numerical results to the observations, respectively.  Assuming that the density
enhancement in the near zones of the fgQSOs is the same as that obtained from
the LOSPE, the TPE has been simulated for both type 1 and type 2 fgQSOs with a
wide range distribution of the fgQSO properties, i.e., the lifetime and the
half opening angle of the associated tori. Our numerical simulations show that
the response of the TPE to the change in the fgQSO lifetime (or the luminosity
evolution or fluctuation on timescale of $10^6-10^7$~yr) and that to the change
of the half opening angle of torus are different. The differences in the
resulted TPE for different settings of the QSO lifetime are large for the
high-luminosity samples, but they are relatively small for the low-luminosity
sample.  These results suggest that the density enhancement near the fgQSOs,
the fgQSO lifetime and the half opening angle of the tori associated with the
fgQSOs can be simultaneously constrained by the LOSPE and the TPE combined
together of a sample with several hundreds bright QSOs.
If a significant number of type 2 QSOs can be revealed by future observations,
our simulations show that the TPE of type 2 QSOs can be significantly different
from that of type 1 QSOs, and an observational search for the contrast will
further help to distinguish the different effects due to QSO properties or
environment and improve the constraints.

The lifetime and the opening angle of the torus are of fundamental importance
for our understanding of the QSO physics and the growth of massive black holes
(MBHs). The lifetime of QSOs is an important parameter characterizing the
luminosity evolution of QSOs and thus the assembly history of MBHs (including
the mass and the spin evolution). A number of independent arguments suggest
that the lifetime of bright QSOs is on the order of $10^7-10^8\yr$ (e.g.,
\citealt{HH01,MW01,YT02,Steidel02,YL04,Marconi04,YL08,Shen09,Shankar10}; also
see a review by \citealt{Martini04}). However, in principle, MBH growth may
occur mainly through a long period continuous accretion or alternatively many
short-period accretion episodes; and those numbers are mainly obtained from
demography of QSOs and they may only represent the net lifetime that the QSO
luminosity in some wavelength range larger than a threshold, not the time
period for each accretion episode (e.g., see \citealt{Martini04}).  As we have
demonstrated that the TPE offers a way, if not the only way, to constrain the
episodic lifetime of QSOs \citep[see also][]{Adelberger04}, which should
provide considerable insight into our understanding of the growth history of
MBHs.  The presence of a torus surrounding each QSO is proposed to be the
underlying reason that leads to the classification of type 1 QSOs and type 2
QSOs. The half opening angle of the torus is an important parameter that
determines the ratio of type 2 to type 1 QSOs. As demonstrated in this paper,
the TPE, together with the LOSPE, can provide constraints on this angle, which
is independent of those previous ways mainly through observational ratios of
type 2 to type 1 QSOs \citep[e.g.,][]{TU06,Gilli07,Treister10}. As an
alternative potential tool to constrain the opening angle of the torus, the QSO
PE may be helpful in our understanding of the unification picture of different
types of the AGN population and its underlying physics.

KT08 measured both the LOSPE and the TPE for a low-redshift low-luminosity QSO
sample ($z\sim 2$ and $L_{\nu_0}\sim 5\times 10^{30}\ergshz$), and their
measurement on the TPE suggested that the QSO lifetime is $\la 10^6$~yr. That
is apparently in contradiction with the detections of the TPE through other
observations on He~{\small II} or metal lines, which require a longer QSO
lifetime $\tauq\ga (2-3) \times 10^7$~yr (e.g.,
\citealt{Jakobsen03,WW06,Worseck07,GSP08}). Through our Monte-Carlo
simulations, we have shown that the apparent contradiction can be solved after
considering a combination of the effects due to density enhancement in the near
zone of the fgQSOs and the obscuration of the tori associated with the fgQSOs.

\citet{Guimaraes07} measured the LOSPE for a high-redshift high-luminosity sample
($z\sim 4$ and $L_{\nu_0}\sim 8\times 10^{31}\ergshz$), and their LOSPE measurement 
suggests that the density in the near zones of those QSOs is enhanced. In order to 
reproduce their measurement on the LOSPE, a density enhancement is also required in
our Monte-Carlo simulations. Compared with \citet{Guimaraes07} estimates on the density enhancement, 
the required density enhancement here is smaller by a factor of $1.5-5$ at a distance
of $R_{||}\la 10$~Mpc-$1$~Mpc but similar at a distance of $R_{||}\ga 15$~Mpc.
The main reason for this difference is that the Poisson variance of absorption lines in
the near zones of QSOs, which is included in our Monte-Carlo simulations,
leads to an averaged absorption that is larger than a simple estimation obtained without
considering of the Poisson variance \citep[see][]{Guimaraes07}. Therefore, 
a less significant density enhancement is required here compared with that 
obtained by \citet{Guimaraes07} (see also discussions on the Poisson variance 
in \citealt{Dall08}).

We conclude that the current measurements on the LOSPE and the TPE are consistent
with that (1) the density is significantly enhanced in the vicinity of fgQSOs, (2)
the fgQSO lifetime is longer than a few $10^7$~yr and a much shorter lifetime
(i.e., $\la 10^6$~yr) is excluded or at least is not required, and (3) the half
opening angle of the tori associated with  fgQSOs is $\sim 60\degr  $, which
are consistent with other independent estimates
\citep[e.g.,][]{Jakobsen03,WW06,Worseck07,GSP08,TU06,Gilli07,Treister10}.
Future observations on the PE of several hundreds of bright type 1 and type 2
(fg)QSOs will potentially set accurate constraints simultaneously on these QSO
properties.

\section*{Acknowledgments}
This work was supported in part by the National Natural Science Foundation
of China under No.\ 10973001, 10973017, 11033001, and 
the Bairen program from the National Astronomical Observatories, Chinese
Academy of Sciences.

\appendix

\section{Intersections of the light cone confined by the torus and
the surface confined by the time-delay effect}  \label{app:Appendix}

The light cone confined by the dusty torus associated with the central engine
of a QSO is given by 
\begin{eqnarray}
\lefteqn{X^2+(Y\cos\Psi_0-R_{\parallel}\sin\Psi_0)^2} \nonumber \\
& & -(Y\sin\Psi_0+R_{\parallel}\cos\Psi_0)^2\tan^2\Theta_0=0, 
\label{eq:lightcone}
\end{eqnarray}
where $(X,Y,R_{\parallel})$ are the three-dimensional Cartesian coordinates
with the QSO being the origin, $\Theta_0$ is the half opening angle of the
torus, $\Psi_0$ is the offset of the torus axis from the LOS of the fgQSO (see
Figure~\ref{fig:f1}). The cylinder with a radius of $R_{\perp}$ is defined as
\be
X^2+Y^2=R^2_{\perp},
\label{eq:Rperp}
\ee
where $X=R_{\perp}\sin\Phi$, $Y=R_{\perp}\cos\Phi$, and $\Phi\in [0\degr
,360\degr  )$ is the rotation angle around the LOS. Figure~\ref{fig:f1} shows
the cases for $\Phi=0\degr  $. For any given $\Phi$, the intersections of the
light cone with the cylinder are the solutions of equations \ref{eq:lightcone}
and \ref{eq:Rperp}.  If $\Phi=0\degr  $, the two solutions are
$R_{\parallel,\rm E}=-R_{\perp}/\tan(\Theta_0-\Psi_0)$ and $R_{\parallel,\rm
F}=R_{\perp}/\tan(\Theta_0+\Psi_0)$ for type 1 QSOs (the left panel in Figure\
\ref{fig:f1}) or $R_{\parallel,\rm E}=R_{\perp}/\tan(\Theta_0+\Psi_0)$ and
$R_{\parallel,\rm F}=-R_{\perp}/\tan(\Theta_0-\Psi_0)$ for type 2 QSOs (the
right panel in Figure\ \ref{fig:f1}), respectively. For the case of $\Phi=90\degr
$, the two solutions are $R_{\parallel,\rm
E}=-{R_{\perp}}/({|\cos\Psi_0|\sqrt{\tan^2\Theta_0-\tan^2\Psi_0}})$ and
$R_{\parallel,\rm F}=-R_{\parallel,\rm E}$ if $|\tan\Theta_0|>|\tan\Psi_0|$;
and there is no solution to the above two equations, otherwise. For those cases
without solutions to the above two equations, there is no the PE on the bgQSO
spectrum due to the dusty torus associated with the type 2 fgQSO.

\end{document}